\documentclass[msmath,amssymb,aps,prl, twocolumn, epsfig, showpacs, bibliography, lengthcheck]{revtex4-1}

\usepackage{graphicx}% Include figure files
\usepackage{dcolumn}% Align table columns on decimal point
\usepackage{bm}% bold math
\usepackage{hyperref}% add hypertext capabilities
\usepackage{epstopdf}

\begin{document}
\title{New classes of systematic effects in gas spin co-magnetometers.}% Force line breaks with \\
\author{D. Sheng, A. Kabcenell, M. V. Romalis}
\affiliation{Department of Physics, Princeton University, Princeton, New Jersey 08544, USA}

%\date{\today}

\begin{abstract}

Atomic co-magnetometers are widely used in precision measurements searching for spin interactions beyond the Standard Model. We describe a new $^3$He-$^{129}$Xe co-magnetometer probed by Rb atoms and use it to identify two general classes of systematic effects in gas co-magnetometers, one associated with diffusion in second-order magnetic field gradients and another due to temperature gradients. We also develop a general and practical approach for calculating spin relaxation and frequency shifts due to arbitrary magnetic field gradients and confirm it experimentally.

\end{abstract}

\maketitle

Spin-dependent interactions are used in many low-energy experiments searching for new physics. Such experiments are often limited by noise and systematic effects associated with the magnetic field, which is the dominant spin interaction in the Standard Model. To reduce these effects, a number of experiments rely on co-magnetometers, first introduced in \cite{Fortson}, that use two different spin species to measure magnetic fields in the same space and time. Examples of such experiments include searches for EDM of the neutron \cite{nedm} and atoms ~\cite{ChuppEDM,TlEDM}, searches for violation of local Lorentz invariance \cite{Walsworth,brown10,smiciklas11} and for new spin-dependent forces  \cite{venema92,vasilakis,bulatowicz,tullney}.  Co-magnetometers also find practical applications in inertial rotation sensing ~\cite{kornack,Larsen}.

The primary purpose of a co-magnetometer is to use two spin species to measure the same average magnetic field independent of its spatial profile, they usually rely on fast atomic diffusion so that all atoms uniformly sample the measurement volume. It is natural, therefore, to consider effects that limit this cancelation. Some effects of this type have been discussed in connection with neutron EDM experiments \cite{Harris,Lamoreaux,Beck} for specific cases. However, we are not aware of a general analysis in the gas diffusion regime.

In this Letter we use a $^3$He-$^{129}$Xe co-magnetometer probed by Rb atoms to experimentally study the effects of magnetic field gradients and temperature gradients. We find that second-order magnetic field gradients cause shifts in the ratio of the $^3$He and $^{129}$Xe precession frequencies proportional to the third power of the gradient strength. We develop a new approach for theoretical analysis of spin relaxation and frequency shifts due to arbitrary magnetic field gradients and surface spin relaxation. Unlike previous methods~\cite{Cates,Golub,Golub1}, it does not rely on second-order perturbation theory and therefore can describe effects proportional to higher powers of the gradient strength. We expand the spin polarization in diffusion eigenmodes of the Torrey equation \cite{Torrey}, calculate the coupling matrix between the eigenmodes, and find its eigenvalues after truncating very high order modes suppressed by diffusion. This approach works for arbitrary relative size of diffusion, gradient dephasing, and Larmor precession timescales as long as motion of atoms is well described by the diffusion equation.

We also describe the effect of thermal diffusion~\cite{Grew} on co-magnetometers, which was not considered before to our knowledge. It  causes a gradient in the relative concentration of the two spin species in the presence of a temperature gradient and results in a linear sensitivity of the spin precession frequency ratio to first-order magnetic field gradient in the direction of the temperature gradient.

Co-magnetometers using $^3$He and $^{129}$Xe, first introduced in \cite{Chupp}, are a natural choice for precision measurements because both species have nuclear spin $I=1/2$ and long spin coherence times. Previous experiments used inductive pick-up coils \cite{Walsworth,ChuppEDM}  or SQUIDs \cite{tullney} to detect the dipolar magnetic field created by polarized  $^3$He and $^{129}$Xe atoms outside of the cell. We use Rb atoms in the same cell to detect nuclear spins through their Fermi-contact interactions, which enhances the dipolar magnetic field by a factor of 5.6 for $^3$He \cite{Romalis98} and 490 for $^{129}$Xe \cite{Saam}.  However, spin interactions also cause shifts of the nuclear spin precession frequencies due to the Rb polarization \cite{Schaefer}. Therefore, our measurement procedure is designed to suppress the polarization of Rb during the free precession measurement interval for nuclear spin. During this interval we turn off the lasers interacting with Rb atoms and apply a strong oscillating magnetic field at the Rb Zeeman resonance frequency to suppress Rb polarization \cite{Grossetete} generated by spin-exchange with $^{129}$Xe \cite{Baranga}.

\begin{figure}
\includegraphics[width=3in]{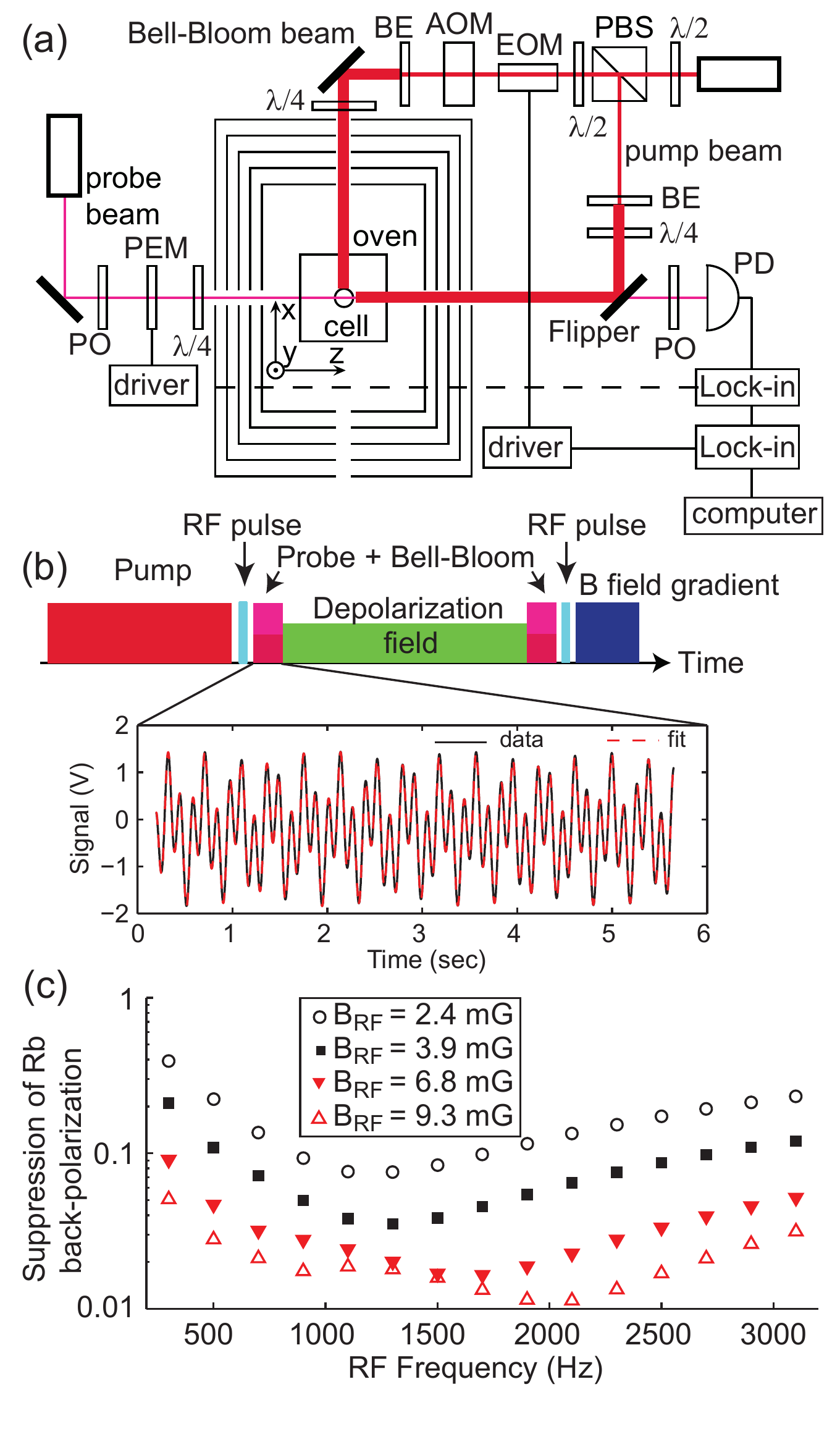}
\caption{\label{fig:setup}(Color online) (a) Top view of experimental setup. BE: beam expander, PBS: polarizing beam splitter,  PO: polarizer, PD: Photodiode, PEM: photo-elastic modulator. (b) The sequence of the pulsed operation and an example of the signal recorded during the measurement pulse. (c) The fractional suppression of the Rb back-polarization generated by spin-exchange with $^{129}$Xe as a function of the amplitude and frequency of the depolarizing field. }
\end{figure}

The measurements are performed in a spherical 1.88~cm internal diameter cell made from GE180 aluminosilicate glass with 3.2 atm $^3$He, 2.9 torr $^{129}$Xe, 70 torr N$_2$, and a droplet of natural abundance Rb with a small admixture of K. The droplet is used to plug the cell stem at its opening to prevent gas diffusion into the stem and improve cell sphericity. The surface spin relaxation time is about 50 sec for $^{129}$Xe and much longer for $^3$He. The cell is placed in a five-layer $\mu$-metal shield and is supported by a G10 rod attached to a 3-axis translation stage outside of the shield to control the cell position relative to the magnetic field and gradient coils mounted inside the shield. The gradient coils are calibrated by measuring the frequency of nuclear spin precession as a function of cell position. A uniform bias field of 2.4~mG generated by a stable current source is applied in $\hat{z}$ direction throughout the measurements. The cell is heated in a boron nitride oven to about 125$^{\circ}$C by AC electric currents, a separate stem heater is used to control the stem temperature independently. We monitor the temperatures of the cell stem, the bottom of the cell body and the oven body with T-type thermocouples. The experimental setup is shown in Fig.~\ref{fig:setup}(a).

The measurements are started by optically pumping Rb atoms along the bias magnetic field for twenty minutes to build up $^3$He polarization. Then the pump beam is blocked and an rf pulse is applied in $\hat{y}$ direction to tip both $^3$He and $^{129}$Xe spins by $\pi/2$. The frequency of nuclear spin precession is determined from two measurement pulses separated by a time $T$ (30$\sim$50 sec) on the order of Xe $T_2$~\cite{sheng13}.  Each measurement pulse uses a pump beam along the $\hat{x}$ direction, whose polarization is modulated between left and right circular at 200 Hz with an electro-optic modulator, and a linearly polarized probe beam along the $\hat{z}$ direction. We measure the polarization rotation of the probe beam caused by the tilt of the total magnetic field experienced by Rb atoms due to the precessing transverse magnetization of the nuclear spins. This measurement procedure minimizes Rb polarization along the bias field. The probe optical rotation signal is demodulated at 200 Hz by a lock-in amplifier before being recorded.  In between the two measurement pulses all lasers are blocked and we turn on an rf field along $\hat{x}$ direction with an amplitude of 4.8~mG and frequency of 1.5~kHz to depolarize Rb spins by saturating their Zeeman resonance. After the second measurement pulse, we recycle the $^3$He polarization by applying a $\pi/2$ pulse with an appropriate phase to put $^3$He spins back along the bias field. It follows by a pulse of magnetic field gradient to relax all transverse nuclear spin components and one minute of optical pumping to build-up the $^{129}$Xe polarization before the next measurement cycle.   Fig.~\ref{fig:setup}(b) shows the measurement sequence and an example of the signal recorded during the measurement pulse.

We fit optical rotation data using the equation
\begin{eqnarray}
\theta&=&A_{\rm Xe}e^{-\frac{t-t_m}{\tau_{\rm Xe}}}\sin[\omega_{\rm Xe}(t-t_{0,\rm Xe})]+b(t-t_m)\nonumber\\
&+&A_{\rm He}e^{-\frac{t-t_m}{\tau_{\rm He}}}\sin[\omega_{\rm He}(t-t_{0,\rm He})]+c,
\end{eqnarray}
where $t_m$ is the center of the measurement pulse, $t_{0,\rm Xe}$ ($t_{0,\rm He}$) are the crossing-zero times of Xe (He). In each measurement pulse, lasting about 5 sec, there are several crossing-zero time points. We choose $t_0$ in the first pulse to be closest to the end of the pulse, and $t_0$ in the second pulse to be closest to the beginning of the pulse. The measurement pulses are also adjusted to turn them on and off close to the crossing-zero times for Xe. In this way, we minimize perturbation in the measurement on the Larmor frequency for Xe, which is more sensitive to the presence of polarized Rb atoms. We find the Larmor frequency from $\omega=2{\pi}N/\Delta{t_0}$, where $\Delta{t_0}$ is the difference of the $t_0$ for the two measurement pulses, and $N$ is an integer number of precession periods in between. The transverse spin relaxation times $\tau_{\rm Xe}$ and $\tau_{\rm He}$ are determined from the ratio of the signal amplitudes in the two pulses. The ratio between $^3$He and  $^{129}$Xe Larmor frequencies is
\begin{equation}\label{eq:fratio}
g=\frac{\omega_{\text{He}}}{\omega_{\text{Xe}}}=\frac{\gamma_{\text{He}}B+\Omega_{E}}{\gamma_{\text{Xe}}B+\Omega_{E}},
\end{equation}
where $\gamma$ is the nuclear gyromagnetic ratio, and $\Omega_{E}$ is the projection of the Earth's rotation onto the bias magnetic field direction in the lab frame.
 
In Fig.~\ref{fig:setup}(c) we show the dependence of the fractional suppression of the Rb back-polarization by $^{129}$Xe on the frequency and amplitude of the Zeeman rf field. The presence of the depolarization field during the measurement interval slightly changes the gyromagnetic ratios, $\gamma'=\gamma_0 J_0(\gamma_0 B_d/\omega_d)$~\cite{Haroche}, where $J_0$ is the zero-order Bessel function, $\gamma_0$ is the unperturbed gyromagnetic ratio, $B_d$ and $\omega_d$ is the magnetic field amplitude and frequency of the depolarization field. This modification is different for  $^3$He and  $^{129}$Xe, and introduces a constant change in the frequency ratio, $g'=g-6.3\times10^{-5}$. It is important that the depolarizing field does not have a rotating component, which introduces a much larger frequency shift. In a separate experiment we investigated the use of a depolarization field at the hyperfine resonance frequency of isotopically enriched $^{39}$K atoms. It produces similar suppression of electron polarization without a significant effect on nuclear spin precession frequencies.

Frequency shifts due to linear magnetic field gradients were first considered in \cite{Cates} using second-order perturbation theory. Their analysis shows that the frequency ratio $g$ does not change up to second order in the gradient strength if the Larmor frequency is much faster than the diffusion time across the cell, $\omega \gg D/R^2$, where $D$ is the diffusion constant and $R$ is the cell radius. This condition is well satisfied in our experiment. However, in practical experiments the field gradients are usually dominated by higher-order terms, either due to field coils or local magnetic impurities. To analyze higher-order gradients and higher powers of gradient strengths we developed a new method for calculating frequency shifts and relaxation rates in the gas diffusion regime.  We start with Torrey equation \cite{Torrey} for the magnetization vector $\mathbf{M}$,
\begin{equation}
\frac{\partial \mathbf{M}}{\partial t}=\gamma \mathbf{M}\times\mathbf{B}+D \nabla^2 \mathbf{M}.
\label{Torrey}
\end{equation}
In a spherical cell the magnetization is expanded in vector spherical harmonics and spherical Bessel functions, while the
magnetic field is expanded in vector spherical harmonics, assuming no magnetic field sources inside the cell:
\begin{eqnarray}
\mathbf{M}(\mathbf{r},t)&=&\sum_{nljm}M_{nljm}(t) \mathbf{Y}^l_{jm}(\theta,\phi) j_l(k_{ln} r/R) \\
\mathbf{B}(\mathbf{r})&=&\sum_{lm} B_{lm} (\sqrt{4\pi l}/l!) r^{l-1} \mathbf{Y}^{l-1}_{lm}(\theta,\phi).
\end{eqnarray}
Equation (\ref{Torrey}) is then converted to a system of linear differential equations for $M_{nljm}(t)$ using orthogonality and completeness of vector spherical harmonics and spherical Bessel functions. The equations are truncated to a maximum order in $l$ and $n$ since diffusion damps out higher order terms. The eigenvalues of the resulting matrix give the decay rates and frequencies of the normal diffusion modes. We evaluate the matrix symbolically as described in the supplemental material \cite{suppl} and then find the eigenvalues numerically for a given diffusion constant and magnetic field specified by $B_{lm}$.  We verified that this approach reproduces all results in \cite{Cates}. It also remains valid when the rate of gradient dephasing is larger than the rate of diffusion across the cell, $\gamma \nabla B R > D/R^2$, where the perturbation theory in \cite{Cates} breaks down.  Spin relaxation on cell walls can also be easily incorporated by modifying the wall boundary conditions used to determine the diffusion mode constants $k_{ln}$, as described in \cite{suppl}. It can be shown, however, that in a spherical cell the frequency ratio $g$ is not affected by the surface spin relaxation if it is isotropic, because the average magnetic field over any spherical shell is equal to the field at the center of the shell \cite{Jackson}. Anisotropic surface relaxation and other cell geometries can be considered using an extension of this approach.

\begin{figure}
\includegraphics[width=3.4in]{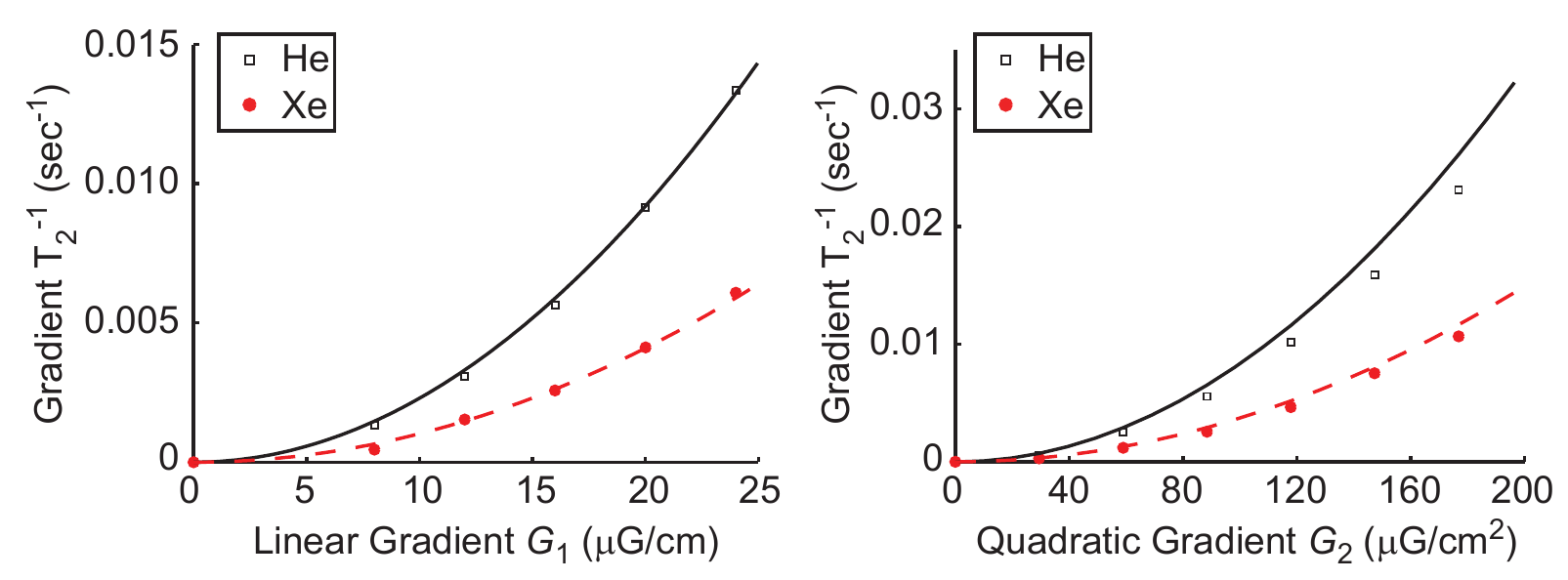}
\caption{\label{relax}(Color online) Measurements of the transverse relaxation rate for $^{3}$He and $^{129}$Xe due to linear (left panel) and quadratic (right panel) magnetic field gradients (points) together with results of the matrix analysis (lines) with no adjustable parameters.}
\end{figure}
To experimentally test our approach we measured the transverse spin relaxation rate due to first and second order gradients, $\delta B_z^{(1)}= G_1 z$ and $\delta B_z^{(2)} = G_2 (z^2/2-x^2/4-y^2/4)$. In Fig.~\ref{relax} we plot the difference between $\tau_{\rm He}^{-1}$ ( $\tau_{\rm Xe}^{-1}$) with and without the applied gradient.  In our cell the diffusion constant for Xe is dominated by binary diffusion in He, while for He it is dominated by the self-diffusion constant, both are inversely proportional to the pressure of He. Using data from \cite{Kestin} we determine the ratio of the two diffusion constants $D_{^{3}He}/D_{Xe-{^3}He}=3.38$ at 125$^{\circ}$C, after correcting for the isotopic mass difference between $^{3}$He and $^{4}$He. We check the initial filling buffer gas  pressure in the cell by measuring the pressure broadening of the Rb D$_1$line, which was recently calibrated in \cite{Todd}. After correcting for the presence of N$_2$, we find  $D_{^{3}He}=0.64$ cm$^2$/sec for our temperature and pressure based on data in \cite{Kestin}. There are no adjustable parameters in the comparison with the model in Fig.~\ref{relax}. To further verify our approach, we also extended the treatment in \cite{Cates} to calculate the transverse relaxation due a longitudinal gradient of order $l$, $\partial^l B_z/\partial z^l$, and find it agrees with our approach for high $l$ \cite{suppl}. For $l=2$ we find $1/T_2=11 \gamma^2 R^6 (\partial^2 B_z/\partial z^2)^2/5880 D$.

\begin{figure}
\includegraphics[width=2.8in]{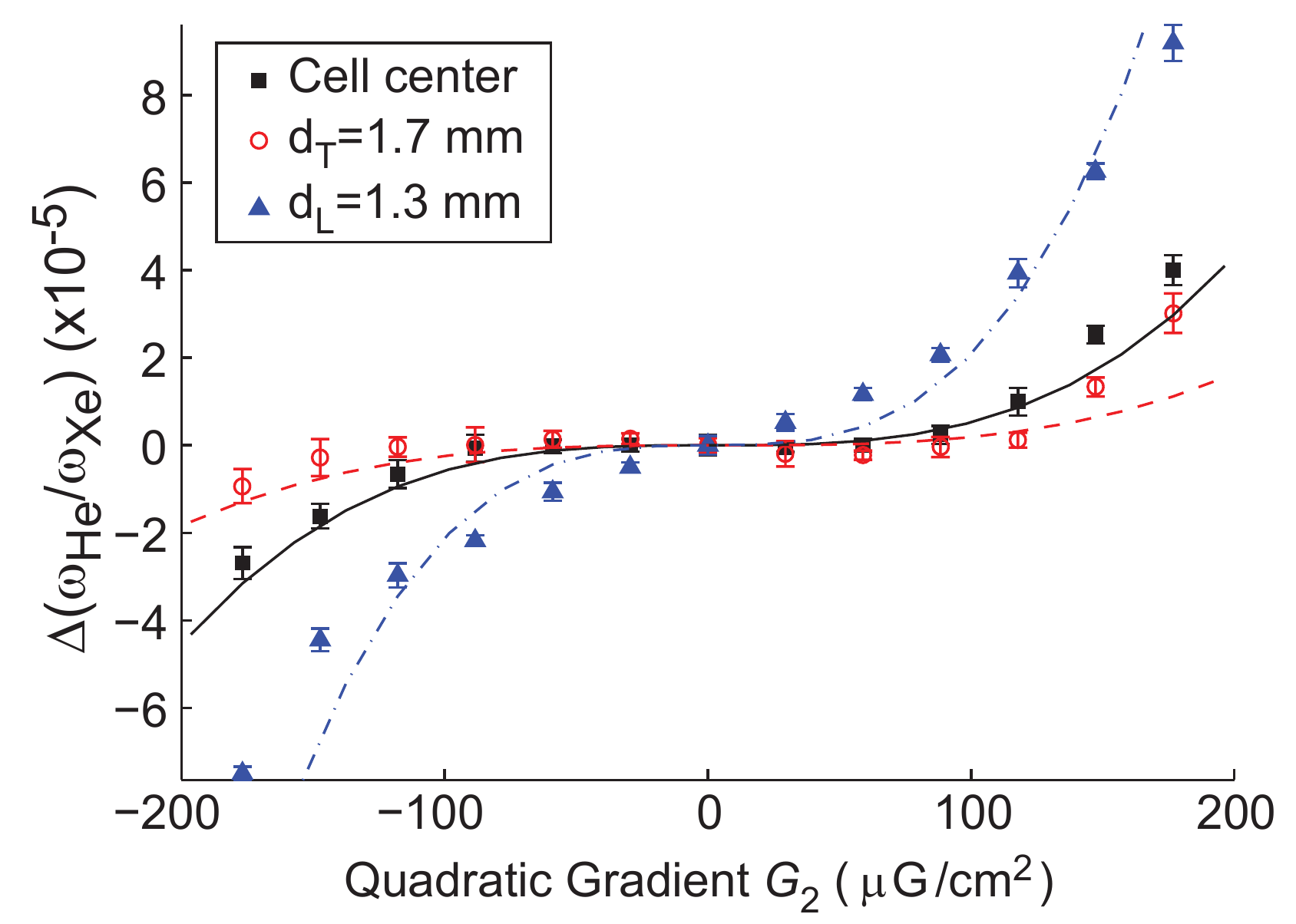}
\caption{\label{shift}(Color online) Changes in the ratio of the spin precession frequencies $g$ due to the quadratic magnetic field gradient $\delta B_z^{(2)}$. Three data are for the cell centered relative to the gradient coil (squares), displaced in the transverse direction by 1.7 mm (empty circles) and displaced along $\hat{z}$ direction by 1.3 mm (triangles). The lines are model predictions with no free parameters.}
\end{figure}

In Fig.~\ref{shift} we show the shifts in the frequency ratio $g$ due to the second order gradient $G_2$.  We find that the second order gradient causes a shift in $g$ proportional to the third power of the gradient strength. This effect is due to non-uniform polarization of $^{3}$He and $^{129}$Xe spins caused by spatial variation in the gradient relaxation rate resulting in a shift of the ``center-of-spin''.  It cannot be described by second-order perturbation theory approaches and can cause systematic effects in precision measurements since it is odd in the gradient sign. It is also very sensitive to the position of the cell relative to the center of the gradient, as illustrated in Fig.~\ref{shift}. From our model and dimensional arguments we find that the shift in $g$ is proportional to the ratio $(\gamma_{He}^2/D_{^{3}He}^2-\gamma_{Xe}^2/D_{Xe-{^3}He}^2)$ and is actually significantly suppressed in our case because the ratio of the diffusion constants is close to $\gamma_{He}/\gamma_{Xe}=2.75$.

\begin{figure}
\includegraphics[width=2.8in]{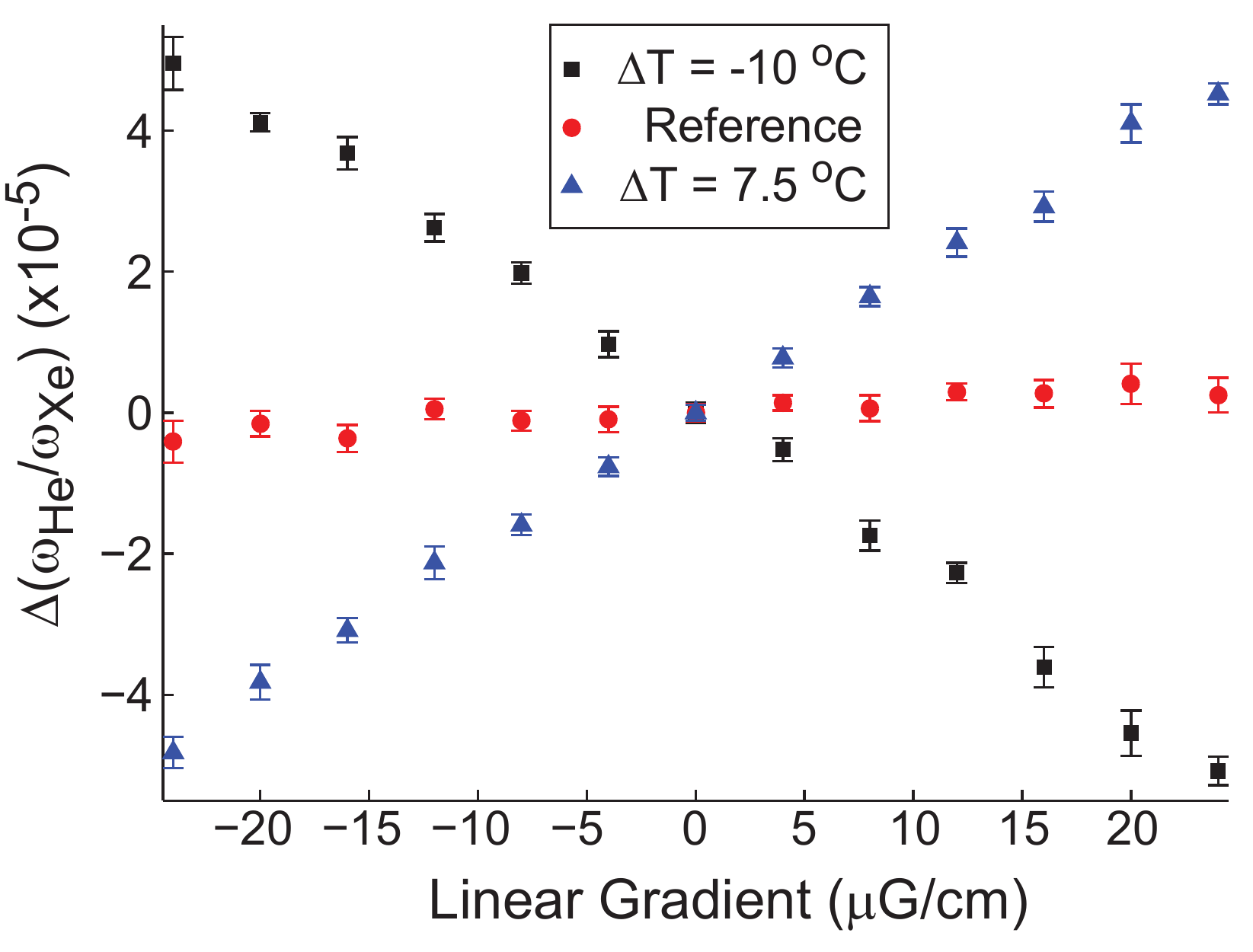}
\caption{\label{tdiff}(Color online) Changes in the response of the frequency ratio $g$ to a linear magnetic field gradient $\partial B_z/\partial y$ in the presence of temperature gradient $\partial T/\partial y$ across the cell. $\Delta T$ denotes the temperature difference between the top and bottom of the cell.}
\end{figure}

When studying the response of the frequency ratio $g$ to first order gradients, we found that it is sensitive to the temperature gradients across the cell. This effect can be attributed to thermal diffusion - a gradient in the relative concentration of $^{3}$He and $^{129}$Xe spins due to a temperature gradient. The relative concentration gradient due to thermal diffusion is given by \cite{Grew}
\begin{equation}
\frac{d f_{Xe}}{d r}=-\alpha_T f_{Xe} f_{He} \frac{1}{T} \frac{d T}{d r},
\end{equation}
where the relative concentrations are $f_{Xe}=n_{Xe}/(n_{He}+n_{Xe})$,  $f_{He}=n_{He}/(n_{He}+n_{Xe}) \approx 1$, and $\alpha_T$ is the thermal diffusion factor. A temperature gradient across the cell causes a non-uniform distribution of both helium and xenon simply to maintain constant pressure, but for xenon the concentration is further enhanced in colder region due to thermal diffusion. This causes a separation of the center-of-spin for the two species and a shift in the frequency ratio $g$ for a linear magnetic field gradient parallel to the temperature gradient. In a spherical cell in a uniform temperature gradient the separation of the centers-of-spin is given by $d=\alpha_T R \Delta T/10 T$, where $\Delta T$ is the temperature difference across the cell. The thermal diffusion coefficient for small  concentration of Xe in He is $\alpha_T=1.06$ \cite{Kestin} and we calculate that for our conditions $d=2.5 \times 10^{-4} \Delta T$ cm/K. Fig.~\ref{tdiff} shows the experimental measurements of the changes in $g$ due to a  vertical linear magnetic field gradient $\partial B_z/\partial y$ for different vertical temperature gradients. We indeed find that the sign of the shift changes with the sign of the temperature gradient and consistent with the sign of thermal diffusion. From these data we find $d=2.0 \times 10^{-4} \Delta T$ cm/K, which is in good agreement with the calculation given the uncertainty in the temperature gradient of the gas inside the cell.

In conclusion, we have described two new general classes of systematic effects that affect gas co-magnetometers. One frequency shift is due to higher-order magnetic field gradients, which have not been investigated before experimentally or theoretically. We developed a high-order method to calculate the effects of field gradients and find that the frequency shift is proportional to the third power of the gradient strength. The second source of frequency shift is due to thermal diffusion effect, which causes gradients in the relative concentration in the gases in the cell in the presence of a temperature gradient, resulting in sensitivity to first-order magnetic field gradients. Identification of these systematic effects will be important for future precision measurements using co-magnetometers, in particular, in searches for spin-gravity coupling \cite{Pospelov} and other interactions where signal reversal is difficult, as well as for their practical applications. This work was supported by NSF and DARPA.


\begin{thebibliography}{99}
\bibitem{Fortson} S. K. Lamoreaux, J. P. Jacobs, B. R. Heckel, F. J. Raab, and E. N. Fortson, Phys. Rev. Lett. {\bf 57}, 3125 (1986).
\bibitem{nedm} C. A. Baker {\it et al.}, Phys. Rev. Lett. {\bf 97}, 131801 (2006).
\bibitem{ChuppEDM}  M. A. Rosenberry and T. E. Chupp, Phys. Rev. Lett. {\bf 86}, 22 (2001).
\bibitem{TlEDM} B. C. Regan, E. D. Commins, C. J. Schmidt, and D. DeMille, Phys. Rev. Lett. {\bf 88}, 071805 (2002).
\bibitem{Walsworth} D. Bear, R. E. Stoner, R. L. Walsworth, V. A. Kosteleck\'{y}, and C. D. Lane, Phys. Rev. Lett. {\bf 85}, 5038 (2000).
\bibitem{brown10} J. M. Brown, S. J. Smullin, T. W. Kornack, and M. V. Romalis, Phys. Rev. Lett. {\bf 105}, 151604 (2010).
\bibitem{smiciklas11} M. Smiciklas, J. M. Brown, L. W. Cheuk, S. J. Smullin, and M. V. Romalis, Phys. Rev. Lett. {\bf 107}, 171604 (2011).
\bibitem{venema92} B. J. Venema, P. K. Majumder, S. K. Lamoreaux, B. R. Heckel, and E. N. Fortson, Phys. Rev. Lett. {\bf 68}, 135 (1992).
\bibitem{vasilakis} G. Vasilakis, J. M. Brown, T. W. Kornack, and M. V. Romalis, Phys. Rev. Lett. {\bf 103}, 261801 (2009).
\bibitem{bulatowicz} M. Bulatowicz {\it et al.}, Phys. Rev. Lett. {\bf 111}, 102001 (2013).
\bibitem{tullney} K. Tullney {\it et al.}, Phys. Rev. Lett. {\bf 111}, 100801 (2013).
\bibitem{kornack} T. W. Kornack, R. K. Ghosh, and M. V. Romalis, Phys. Rev. Lett. {\bf 95}, 230801 (2005).
\bibitem{Larsen} M. Larsen, M. Bulatowicz, Proc. IEEE Int. Freq. Contr. Symp., Baltimore (2012).
\bibitem{Harris} P. G. Harris, J. M. Pendlebury, and N. E. Devenish, Phys. Rev. D {\bf 89}, 016011 (2014).
\bibitem{Lamoreaux} S. K. Lamoreaux, Europhys. Lett. {\bf 58},  718 (2002).
\bibitem{Beck} G. Baym, D. H. Beck, and C. J. Pethick, Phys. Rev. B {\bf 88}, 014512 (2013).
\bibitem{Cates} G. D. Cates, S. R. Schaefer, and W. Happer, Phys. Rev. A {\bf 37}, 2877 (1988).
\bibitem{Golub} R. Golub, Ryan M. Rohm, and C. M. Swank, Phys. Rev. A {\bf 83}, 023402 (2011).
\bibitem{Golub1} R. Golub, A. Steyerl,  arXiv:1403.0871; M. Guigue, R. Golub, G. Pignol, A. K. Petukhov, arXiv:1403.5530.
\bibitem{Torrey} H. C. Torrey, Phys. Rev. {\bf 104}, 563  (1956).
\bibitem{Grew} K. E. Grew and T. L. Ibbs, {\it Thermal diffusion in gases}, Cambridge Univ. Press (1952).
\bibitem{Chupp} T. E. Chupp, E. R. Oteiza, J. M. Richardson, and T. R. White, Phys. Rev. A {\bf 38}, 3998 (1988).
\bibitem{Romalis98} M. V. Romalis and G. D. Cates, Phys. Rev. A {\bf 58}, 3004 (1998).
\bibitem{Saam} Z. L. Ma, E. G. Sorte, and B. Saam, Phys. Rev. Lett. {\bf 106}, 193005 (2011).
\bibitem{Schaefer} S. R. Schaefer, G. D. Cates, T.-R. Chien, D. Gonatas, W. Happer, and T. G. Walker, Phys. Rev. A {\bf 39}, 5613 (1989).
\bibitem{Grossetete} F. Grossetete, Compt. Rend. {\bf 258}, 3668 (1964).
\bibitem{Baranga} A. B. Baranga, S. Appelt, M. V. Romalis, C. J. Erickson, A. R. Young, G. D. Cates, and W. Happer, Phys. Rev. Lett. {\bf 80}, 2801 (1998).
\bibitem{sheng13} D. Sheng, S. Li, N. Dural, and M. V. Romalis, Phys. Rev. Lett. {\bf 110}, 160802 (2013).
\bibitem{Haroche} S. Haroche, C. Cohen-Tannoudji, C. Audoin, and J. P. Schermann, Phys. Rev. Lett. {\bf 24}, 861 (1970).
\bibitem{suppl} Supplemental material available ...
\bibitem{Jackson} J. D. Jackson, {\it Classical Electrodynamics}, 2nd Ed. Eq. (5.63), Wiley (1975).
\bibitem{Kestin}  J. Kestin, K. Knierim, E. A. Mason, B. Najafi, S. T. Ro and M. Waldman, J. Phys. Chem. Ref. Data {\bf 13}, 229 (1984).
\bibitem{Todd} K. A. Kluttz, T. D. Averett, and B. A. Wolin, Phys. Rev. A {\bf 87}, 032516 (2013).
\bibitem{Pospelov} V. Flambaum, S. Lambert, and M. Pospelov, Phys. Rev. D {\bf 80}, 105021 (2009).
\end{thebibliography}
\end{document}